\documentclass[showpacs,preprintnumbers,showkeys,onecolumn,prl,floatfix]{revtex4}
\usepackage{epsfig}

\begin{document}

\title{Scaling in Plasticity-Induced Cell-Boundary Microstructure:
Fragmentation and Rotational Diffusion.}
\date{\today}

\author{James P. Sethna, Valerie R. Coffman}
\affiliation{Laboratory of Atomic and Solid State Physics (LASSP), Clark Hall,
Cornell University, Ithaca, NY 14853-2501, USA}
\author{Eugene Demler}
\affiliation{Physics Department, Lyman Labs, Harvard University, 
Cambridge, MA 02138}
\date{\today}

\begin{abstract}
We develop a simple computational model for cell boundary evolution in
plastic deformation. We study the cell boundary size distribution and
cell boundary misorientation distribution that experimentally have been
found to have scaling forms that are largely material independent. The
cell division acts as a source term in the misorientation distribution
which significantly alters the scaling form, giving it a linear slope at
small misorientation angles as observed in the experiments. We compare
the results of our simulation to two closely related exactly solvable
models which exhibit scaling behavior at late times: {\it
(i)}~fragmentation theory and {\it (ii)}~a random walk in rotation space
with a source term. We find that the scaling exponents in our simulation
agree with those of the theories, and that the scaling collapses obey
the same equations, but that the shape of the scaling functions depend
upon the methods used to measure sizes and to weight averages and
histograms.
\end{abstract}

 
\maketitle

\section{Introduction}
\label{sec:Introduction}

After significant plastic deformation, the dislocation tangles in
crystals often organize themselves into sharp walls separating nearly
dislocation-free cells; the crystallographic axes rotate slightly across
each cell boundary. These cells
undergo refinement (become smaller) under increased deformation, and
recent experiments\cite{HughesPRL,HughesActa} indicate that both the
misorientation angles and the cell sizes have power-law scaling with
material-independent scaling forms for the probability distributions. We
introduce here a simple model of cell division and rotational diffusion
which exhibits this type of scaling, and which provides insights into
the origins for the experimental scaling distributions. In particular,
we argue that cell division (driving the refinement) is responsible for
the linear growth of the misorientation scaling distribution at small
angles.

Cell boundaries are distinct from grain boundaries in that
their misorientation angle across them is small (at most a few degrees)
and they form in a nonequilibrium process, typically at temperatures
where diffusion is not relevant (so, for example, the impurity segregation 
characteristic of many grain boundaries is not observed at cell boundaries).
As deformation proceeds, the cell structure refines (the average cell size
$L_{av}$ becomes smaller), and the average cell misorientation angle 
$\theta_{av}$ grows. 

The cell boundaries are separated into into two classes. An early 
work\cite{Bay} called them ``ordinary cell walls'' and ``dense dislocation
walls''; later authors\cite{HughesPRL} have called them 
GNBs (``geometrically necessary boundaries'') and IDBs 
(``incidental dislocation boundaries''). The GNB's typically align
roughly parallel to one another, have larger misorientation angles, and
are longer, often forming the boundaries of two or more cells. 

(The term ``geometrically necessary'' is unfortunate. Geometrically necessary
{\it dislocations} are those required to mediate macroscopic strain gradients
and rotation gradients, as distinguished from geometrically unnecessary
dislocations whose Burgers vectors cancel out on long length scales. All cell
walls are associated with small relative rotations between cells, and
are hence composed of geometrically necessary dislocations on scales comparable
to the cell sizes. On longer scales, far from building up large macroscopic
rotations, the rotations mediated by neighboring geometrically necessary
{\it boundaries} tend to cancel\cite{HughesPrivateCommunication}, leading 
to little or no long-range rotation gradient. Hence, as cell boundaries, 
the GNB's are most akin to the geometrically unnecessary dislocations.)

Hughes 
{\it et al.}\cite{HughesPRL,HughesActa} studied the distribution functions 
for these two types of cell walls, and found a simple scaling behavior, largely
independent of material. In particular, for the GNBs, 
$\theta_{av} \sim \epsilon^{2/3}$ and $L_{av} \sim \epsilon^{-2/3}$,
while for the IDBs 
$\theta_{av} \sim \epsilon^{1/2}$ and $L_{av} \sim \epsilon^{-1/2}$,
where $\epsilon$ is the magnitude of the
net plastic strain. Moreover, data for several materials and different
strain amplitudes all collapse onto apparently universal scaling curves
$\rho_{\rm mis}$ and $\rho_{\rm size}$ when rescaled to the average angle:
\begin{equation}
\rho(\theta) = \theta_{av}^{-1} \rho_{\rm mis}(\theta/\theta_{av})
\label{eqn:ScalingAngle}
\end{equation}
and
\begin{equation}
\rho(L) = L_{av}^{-1} \rho_{\rm size}(L/L_{av}).
\label{eqn:ScalingLength}
\end{equation}
We will study the scaling of these probability distributions
$\rho(\theta)$ and $\rho(L)$ using a simple model.

\section{Model}
\label{sec:Model}

How much of this apparently universal behavior can be captured in a simple
model of cell rotation and refinement? The model we propose is one in which
cells become smaller by subdivision (leading to a fragmentation theory
for the size distribution), and undergo random angular reorientations as the
strain increases. Our model does not incorporate the anisotropy in the 
external strain field, and so has nothing to say about how cell structure
morphology might change, say, between tensile and rolling deformation or
as the crystalline orientation changes. One should view our model as a 
caricature of the real system; our results suggest that the experimentally
observed scaling behavior may be generic to any microscopic mechanism
which fragments and randomly reorients cells.

Our computational model starts with one large cubical cell. We assume a
cubic crystal, with initial crystalline axes aligned with the axes of our
cube, so the initial orientation is described by a rotation matrix $R(0)$
equal to an identity matrix. The dynamics of our model incorporates
two pieces: rotational diffusion and cell splitting events. 

\noindent
{\bf Rotational Diffusion.} The orientation of each cell $\alpha$ undergoes a
simple random walk in rotation angle space, with strain increments playing the
role of the time step. It is convenient to write
the current orientation $R_\alpha(t) = \exp({\bf n}\cdot{\bf J})$,
where the matrix ${\bf J}_i = \epsilon_{ijk}$ (with $\epsilon_{ijk}$
the totally antisymmetric tensor) generates an infinitesimal rotation 
about the $i^{\rm th}$ axis.  Since the cell boundary
misorientation angles experimentally are small (around a degree or so),
we may expand the exponential in this expression,
\begin{eqnarray}
R =& \exp({\bf n}\cdot{\bf J}) \label{eqn:InfinitesimalRotation}\cr
 \simeq& \left( \begin{array}{ccc}
1-{n_2^2+n_3^2 \over 2} & n_3 & -n_2 \\
-n_3 & 1-{n_1^2+n_3^2\over 2} & n_1 \\
n_2 & -n_1 & 1-{n_1^2 + n_2^2\over 2}.
\end{array} 
\right)
\end{eqnarray}
(Large angle corrections are discussed in \cite{PantleonHansenActa}
and \cite{Dawson}.)
In this approximation, diffusion in the manifold of crystalline orientations
can be written as an ordinary diffusion equation in the three-dimensional
coordinate $\bf n$. If we assume $\epsilon(t)$ is a monotonically increasing
strain, then the three-dimensional probability distribution of grain 
orientations $\Lambda({\bf n})$ evolves
according to the equation
\begin{equation}
\partial \Lambda({\bf n})/\partial \epsilon 
	= -D \nabla^2 \Lambda({\bf n}),
\label{eqn:OrientationalDiffusion}
\end{equation}
where $D$ is the ``orientational diffusion constant'' and the Laplacian
$\nabla^2 = \partial^2/\partial n_x^2 + \partial^2/\partial n_y^2 +
\partial^2/\partial n_z^2.$
The random walk described by this diffusion equation is implemented numerically
by adding a Gaussian random vector to ${\bf n}$ with components of 
root-mean-square length $\sqrt{2 D \Delta \epsilon}$ whenever the strain for
the cell is incremented by $\Delta\epsilon$.

{\bf Cell Splitting Events.} Our model, for simplicity, divides cells
only along planes perpendicular to one of the crystalline axes. Thus our
cell structure is composed of rectangular parallelepipeds. The rate of
cell division in our model depends only on the current size and shape of
the cell, and not on its environment. There are several different
physical mechanisms that might be responsible for cell division. Broadly
speaking, we classify them by dimensional analysis: there are mechanisms
that divide cells at a rate proportional to their current volume $V$,
their current surface area $S$, their current perimeter $P$, or at a
uniform rate independent of the current size $U=1$. Once a cell has been
chosen to split, we must choose an axis and a position along that axis
to place the new cell wall. To keep our aspect ratios reasonable, we've
chosen the probability of splitting along a given axis proportional to
the length of the cell along that axis.  The position of the new cell wall along
the split axis is chosen randomly in all cases. The two cells formed by
splitting inherit their parent's orientation: the new cell walls thus start out
at zero misorientation angle, which will be important when we study the
misorientation angle distribution.

Our model for cell splitting is closely related to a well-studied model
of fragmentation\cite{ZiffMcGrady,ChengRedner}. In fragmentation theory,
the splitting rate is assumed a function of the volume, so for example a
cell of volume $V$ could fracture with a rate $A V^\gamma$. In our
problem, $\gamma=0$ and $\gamma=1$ correspond precisely to uniform and
volume cell splitting rates, while $\gamma=1/3$ and $\gamma=2/3$
approximately correspond to perimeter and area splitting. The evolution
law in fragmentation theory corresponding to our model is easily seen to
be\cite{ChengRedner}
\begin{equation}
\partial c(V,\epsilon)/\partial \epsilon = -A V^\gamma c(V,\epsilon)
+ 2 \int_V^\infty A {\tilde V}^{\gamma-1} c({\tilde V},\epsilon) d{\tilde V}
\label{eqn:FragmentationEvolution}
\end{equation}
where $c$ is the concentration of fragments with volume $V$. To relate
this to experimental measurements (equation~\ref{eqn:ScalingLength}),
which produce probability distributions of lengths rather than
concentrations of volumes, we can change variables from $V$ to
$L=V^{1/3}$. The probability distribution of lengths with all cells weighted
uniformly is
\begin{equation}
\rho(L)=
 {3L^2c(L^3,\epsilon)\over{\int_0^\infty c(V,\epsilon) dV}}.
\label{eqn:cToRho}
\end{equation}

\begin{figure}[thb]
\begin{center}
\epsfxsize=2.5truein
\epsffile{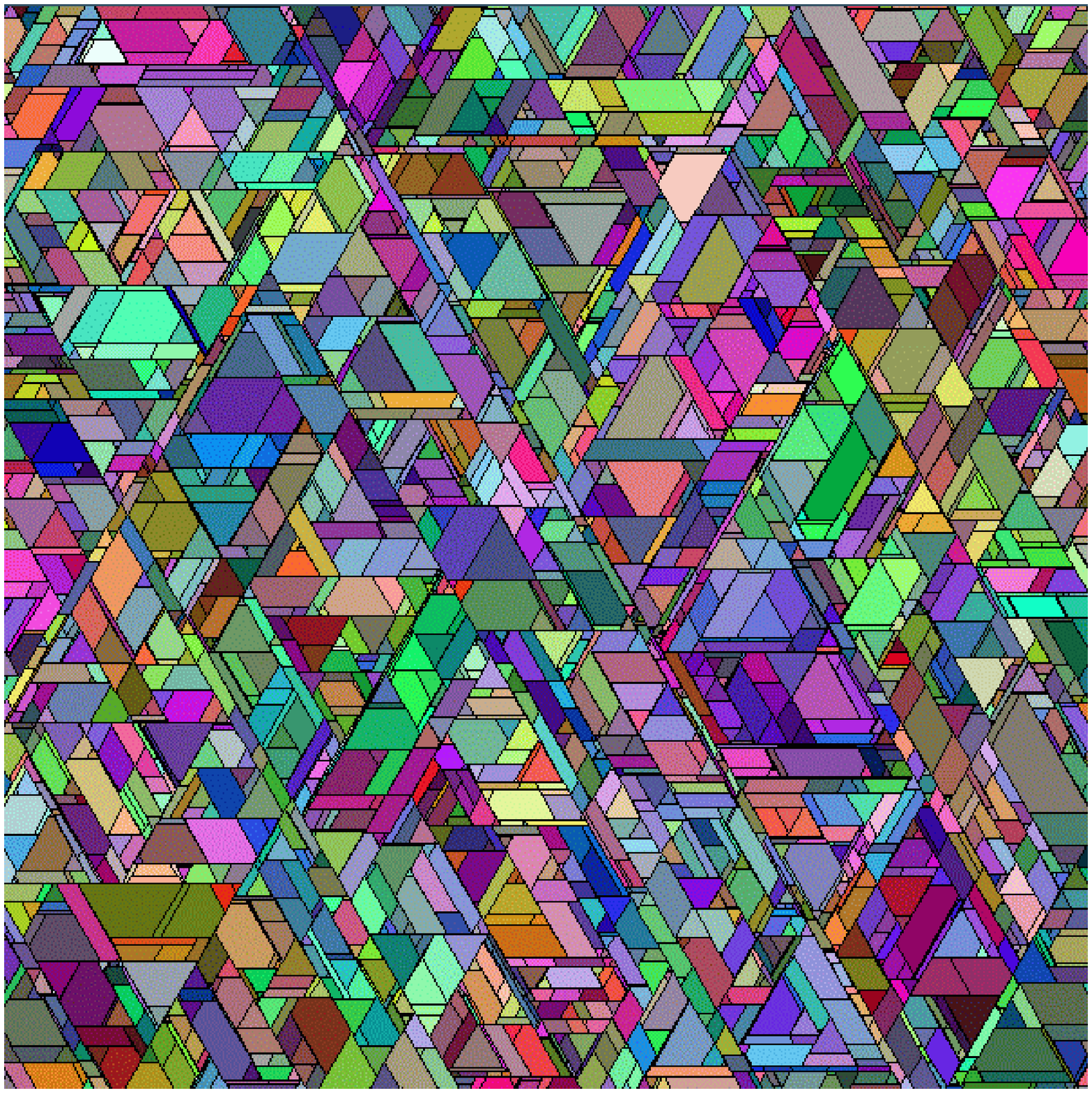}
\hskip 0.1truein
\epsfxsize=2.5truein
\epsffile{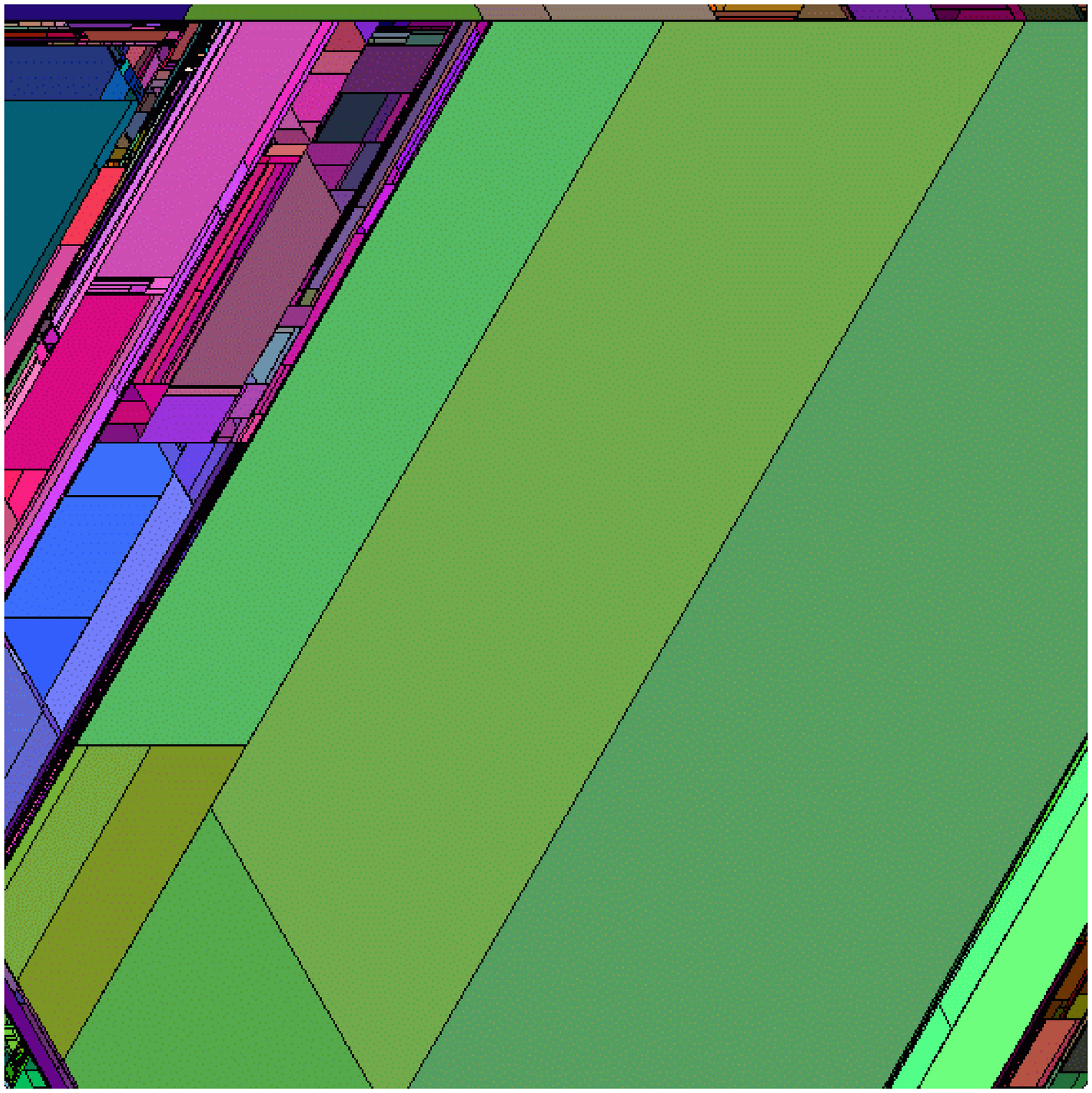}
\end{center}
\caption{{\bf Simulated Cell Morphologies.} Diagonal cross section of
the cell morphologies in two model simulations, in a central plane
perpendicular to the 111 axis. LEFT: Area splitting rate. 
RIGHT: Uniform splitting rate. The area splitting yields rather
uniform cell sizes, while the uniform splitting yields an enormous
range of cell sizes and a fractal morphology. The colors are chosen
to represent the rotations of the crystalline axes of the individual
grains. The original orientation ${\bf n} = 0$ is colored gray; $n_x$,
$n_y$, and $n_z$ are mapped respectively onto deviations in red, green, 
and blue with a scale-factor chosen to saturate at the largest rotations.
}
\label{fig:CellMorphologies}
\end{figure}

The histograms produced by our simulations are a result of three
choices. First, there is a choice in how we define the size (length $L$)
of the cell. In computing the averages and histograms from the
simulation data, we typically 
define the size of a cell to be its length
along any one of the axes: all three lengths are incorporated into the
averages and histograms. This definition of size corresponds to that
used in experiment. Alternatively, in order to compare the histograms
from the simulation to fragmentation theory (which keeps track only of
the volumes, not the shapes, of the cells), we can define the size of
the cell as the cube root of the volume. Second, there is a choice in
the splitting dynamics as discussed above: cells can divide at rates
proportional to their volume, surface area, perimeter, or at a uniform
rate. In fragmentation theory, the rate of splitting is proportional to
$V^\gamma$ as discussed below: thus fragmentation theory is exact for
our simulation with volume splitting ($\gamma=1$) and uniform splitting
rates ($\gamma=0$), but does not directly apply to the perimeter and
surface simulations, whose splitting rates depend upon the shapes of the
cells as well as their volumes. Finally, one must address how to weigh
the contributions of different cells in the probability distribution.
For example, an experiment which measures cell sizes by taking an XY
cross section and then weighting each observed cell equally in the
average is effectively weighting the three dimensional cells by their
extent in the Z direction (roughly weighting each cell by its perimeter
or by $L$). We compute the averages and histograms from the simulation
data by weighting each cell uniformly or by its volume, surface area, or
perimeter. We will soon see that the scaling exponents for the average
size depend on only the splitting dynamics, $\gamma$, not on the
measurement of size or the weight of the distribution. We will also see
that the shape of the scaled probability distributions changes with
different measurements of size or weights, but the distributions scale
nonetheless.

If we define the size of the cell as the cube root of the volume,
simulations that split cells at a uniform rate or at a rate proportional
to the volume of the cell produce histograms that agree well with those
given by fragmentation theory (equation \ref{eqn:cToRho}).  However,
simulations that split cells at rates proportional to area and perimeter
produce histograms that are shifted from equation~(\ref{eqn:cToRho}). We
focus here on the case where the rate is proportional to the surface area $S$
of the cell (Area) and the case where the rate is independent of the
size and shape of the cell (Uniform).

\section{Results}
\label{sec:Results}

\subsection{Morphologies}
\label{sec:Results:Morphologies}

\begin{figure}[thb]
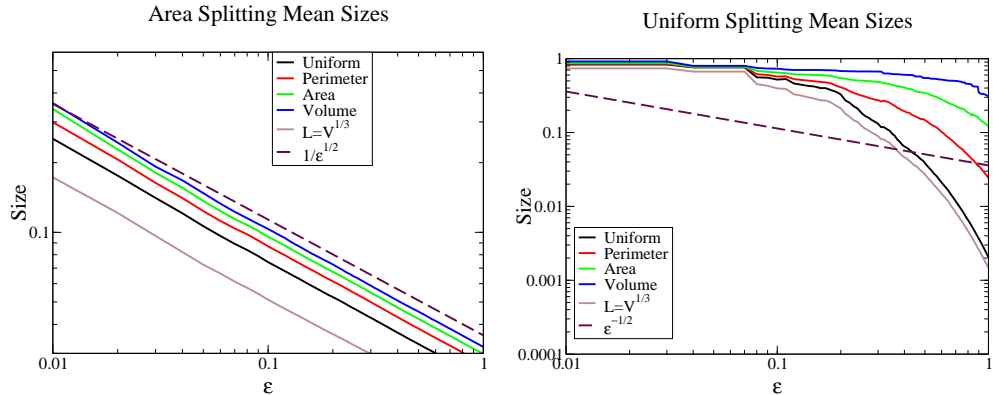

\begin{center}
\epsfxsize=2.5truein
\epsffile{MeanSizesSA.eps}
\hskip 0.1truein
\epsfxsize=2.5truein
\epsffile{MeanSizesRI.eps}
\end{center}
\caption{{\bf Mean Cell Sizes versus Strain.} 
We measure the mean cell sizes, with each cell contributing to the average
an amount given by volume, area, perimeter, or uniformly
independent of its size. As noted in the text, the perimeter scaling
corresponds effectively to a common experimental procedure of taking
histograms. (Don't confuse, say, area weighted mean sizes and area weighted 
splitting rates. The different curves on each graph are different 
{\it measurements}, the different graphs are different {\it dynamics}.)
We also include the mean size with the length measured as the 
cube root of the volume, with cells contributing uniformly, 
independent of their sizes.
LEFT: Area splitting. The various measures are equivalent up
to an overall scale
factor for the model where the cell division rate is proportional to the
surface area of the cells, and the cell size varies as $\epsilon^{-1/2}$.
This equivalence is because the distribution of sizes is peaked about a typical
size scales (figure \protect\ref{fig:CellSize}), and can be derived from the
scaling form for the cell size distribution. RIGHT: Uniform splitting.
The various measures are
very different for the case of random splitting, since the cell structures
are fractal with cells of all sizes.
}
\label{fig:MeanSizes}
\end{figure}

Figures (\ref{fig:CellMorphologies}) show the cell morphologies from the
two simulations. Area splitting shows a fairly uniform density of cell
sizes: this is characteristic also of the other size-dependent cell
division rates. Uniform cell splitting rates produce a broad range of
sizes: most cells are very small, so most cell divisions subdivide very
small cells. Indeed, as we shall discuss below, the uniform model is at
a critical point in fragmentation theory (the ``shattering''
transition\cite{ZiffMcGrady}). Experimentally, there does not seem to be
a consensus on whether the cell size distribution is 
fractal\cite{FractalRefs} or whether it has a more traditional scaling
distribution with a characteristic size which shrinks with
time\cite{HughesActa}.

\subsection{Cell Size Scaling}
\label{sec:Results:Sizes}

Figures (\ref{fig:MeanSizes}) shows the mean cell size as it evolves
with increasing strain in our model, under area splitting and random
splitting. In the figure, we show a count average where each cell contributes
equally, a perimeter average where cells are weighted by their perimeter, and
area and volume average.  Each of these has the size of the cell defined as the
distance between parallel cell walls. The fourth average has the size defined as
the cube root of the volume and is weighted uniformly.  

The mean size for the cells in the Area splitting model scales with
$\epsilon^{-1/2}$. Assuming fragmentation theory
(equation~\ref{eqn:FragmentationEvolution}) and scaling
(equation~\ref{eqn:ScalingLength}, which can be derived from
fragmentation theory\cite{ChengRedner} for $\gamma > 0$), it is
easy\footnote{Write $\bar V = \int \rho_{\rm vol}(V)\, dV$, use equation
\protect\ref{eqn:ScalingLength} to compute $\partial \bar V / \partial
\sigma$, change variables to $y=V/\bar V$, and substitute in the scaling
form equation \protect\ref{eqn:ScalingLength}.} to derive the power-law
relation $L_{av} \sim \epsilon^{-1/3\gamma}$. Thus the experimentally
observed scaling of the IDB sizes suggests a cell splitting rate
proportional to the cell surface area (left side of the figures), while
the scaling $L_{\rm av}^{(GNB)} \sim \epsilon^{-2/3}$ of the GNB
dislocations would suggest a mechanism with a cell splitting rate
scaling as the cell volume to the $1/2$ power. (See, however, part~(2)
in the conclusion.)

\begin{figure}[thb]
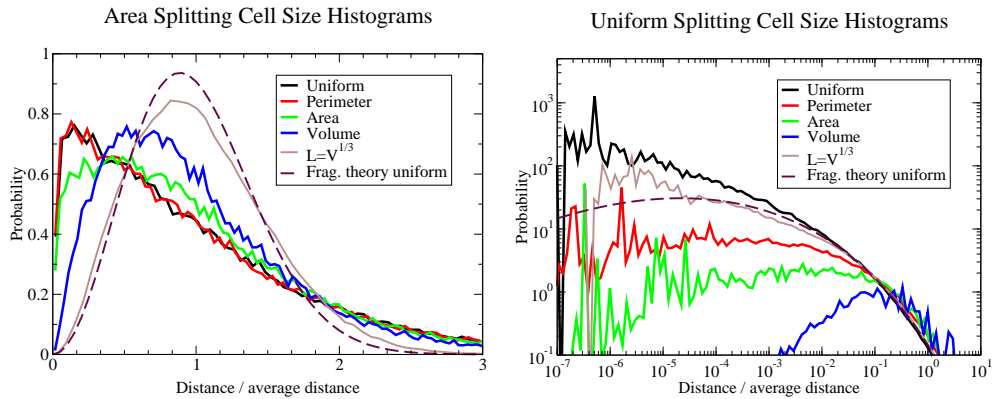

\begin{center}
\epsfxsize=2.5truein
\epsffile{CellSizeSA.eps}
\hskip 0.1truein
\epsfxsize=2.5truein
\epsffile{CellSizeRI.eps}
\end{center}
\caption{{\bf Cell Size histogram.} 
Histograms of cell sizes at the end of our simulation, rescaled to the
average cell size (figure \protect\ref{fig:MeanSizes}) according to
equation (\protect\ref{eqn:ScalingLength}). Again, there are different
histograms depending upon how the cells are weighted in the average.
LEFT: Area splitting. The dashed curve is the fragmentation theory
prediction\protect\cite{ChengRedner} for the closely related
problem where cells split at a rate proportional to volume $V^{2/3}$. 
RIGHT: Uniform splitting. Notice that most of the cells (Counted)
are very small.
}
\label{fig:CellSize}
\end{figure}

Figures (\ref{fig:CellSize}) show the histograms of cell sizes at the
end of our simulation, with the five weights discussed above. Naturally,
for example, if cells are weighted by their volume there is more weight
in the histogram at larger sizes. 

We first consider the left-hand
panel, showing the simulation results with dynamics which split cells
proportional to their surface area. The Area splitting histograms, taken
at different $\epsilon$ during our simulations, do indeed rapidly
converge to those shown on the left panel of figure~(\ref{fig:CellSize})
(hence validating the scaling form equation~(\ref{eqn:ScalingLength})).
However one defines the size $L$ or how one weights the contribution
of each cell, the histograms collapse onto scaling functions.

Fragmentation theory\cite{ChengRedner} calculates the scaling function
explicitly that describes the distribution of sizes at late times. In
particular, for Area splitting ($\gamma=2/3$), the scaling function in
(\ref{eqn:ScalingLength}) for a uniformly weighted distribution is
\begin{equation} \rho_{size}(x) = 32 x^2 \exp(-4 x^2/\pi) /\pi^2
\label{eqn:FragmentationAreaScaling}
\end{equation}
shown as the dashed line on the left in figure (\ref{fig:CellSize}). As
expected, it does indeed agree well with the size distribution where $L$
is defined as the cube root of the volume.  The shift between these two
curves is due to the difference in dynamics: our simulation splits cells
at a rate proportional to the actual area of the cell, not by $V^{2/3}$.
For Volume splitting, with L defined as the cube root of the volume, the
histograms from the simulation agree with those from fragmentation
theory as expected.

The histograms which measure the length, width, and height of each cell
are broader than those that measure the cube root of the volume.  Cells
with large aspect ratios will contribute one or two dimensions that are
smaller than the cube root of volume and one or two that are larger.
Notice that in figure (\ref{fig:CellSize}) the cell size probability
density does not vanish at zero size $L$ except for statistical
weightings that involve the total volume. In our model, a subdivision
occurs with equal probability at all thicknesses, so the probability
density at zero thickness is finite. If the weights of the cells in the
average is by the total volume, of course, then the thin cells
contribute vanishing weight so the histogram goes to zero.

Consider now the right-hand panel of figure (\ref{fig:CellSize}), with
simulation results for dynamics which split cells at a uniform rate
independent of their size. Notice first that the scales are logarithmic:
there is an enormous range of cell sizes, with peak probabilities (by
most measures) at very small sizes. The Uniform splitting model is at a
critical point in the parameter $\gamma=0$, the {\it shattering
transition} \cite{ZiffMcGrady}, beyond which ($\gamma<0$) there an
infinite dust of zero-size particles. At this critical point, there are
exact solutions for the cell size distribution (dashed curve shown in
figure~(\ref{fig:CellSize}), to be compared with the $L=V^{1/3}$ simulation
curve). This exact solution does not have the scaling form of 
equation~(\protect\ref{eqn:ScalingLength}). We have not been able to find a 
generalization of the scaling form which collapses the distribution at the
critical point, but the scaling variable (typical size) must shrink 
exponentially as $\epsilon$ grows. One hint for why finding the scaling
function may be difficult is that the system appears not to be self-averaging:
the relatively good agreement between the theory and simulation shown
on the right of figure~(\ref{fig:CellSize}) is not as true with other
random number seeds, with fluctuations of an order of magnitude away from
the theory. (The average over many seeds does agree with the calculated
form.) While there is no scaling form solution to equation
(\ref{eqn:FragmentationEvolution}) for $\gamma=0$, we have found a family
of formal, non-normalizable solutions of the form 
$c(V,\epsilon)=e^{-(1+2/\beta)\epsilon} V^\beta$.

\subsection{Misorientation Scaling}
\label{sec:Results:Misorientation}

\begin{figure}[thb]
\begin{center}
\leavevmode
\epsfxsize=8.cm
\epsffile{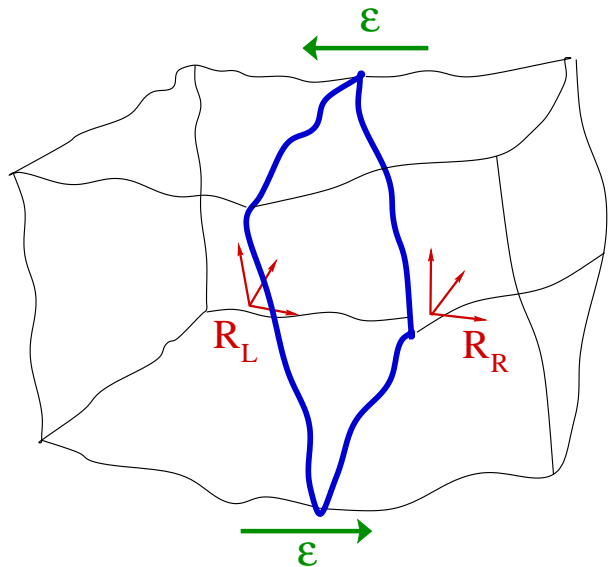}
\end{center}
\caption{{\bf Geometry of a Cell Boundary.} 
}
\label{fig:Geometry}
\end{figure}

To study the misorientation angles, we need a formula for the misorientation
angle $\theta$. Imagine a cell boundary as in figure
(\ref{fig:Geometry}).
The rotation $R_L R_R^{-1}$ takes the crystalline axes of right-hand cell to
the orientation on the left-hand side of the boundary.
As before, we can write this rotation as $\exp({\bf n}\cdot{\bf J})$,
and correspondingly write $R_L$ and $R_R$ in terms of ${\bf n}_L$ and 
${\bf n}_R$. Because the rotation angles are small across cell boundaries,
${\bf n} \simeq {\bf n_L} - {\bf n_R}$. 
The misorientation angle is
\begin{eqnarray}
Tr(R_L R_R^{-1}) =& 1 + 2 \cos \theta \simeq 3 - \theta^2 
= 3 - {\bf n}^2 \cr
\theta \simeq& \sqrt{n_1^2+n_2^2+n_3^2} = |{\bf n}|
\label{eqn:NToTheta}
\end{eqnarray}

Define $\Xi({\bf n}) d^3n$ to be the area of cell boundary with
misorientation matrix $R=\exp({\bf n}\cdot{\bf J})$. Since this distribution
for our model is symmetric under rotations, equation (\ref{eqn:NToTheta})
implies that the probability distribution for the misorientation angle 
\begin{equation}
\rho(\theta) = \rho(|n|) = 4 \pi n^2 \Xi(n) 
\label{eqn:LambdaNToRhoTheta}/A_{tot}
\end{equation} 
where $A_{tot}=\int d^3n \Xi(n)$ is the total cell boundary area.
Notice that the (one-dimensional) probability density for small angles
$\theta$ is $4\pi \theta^2/A_{tot}$ times the (three-dimensional) cell boundary area 
at one of the rotations ${\bf n}$ corresponding to $\theta = |{\bf n}|$. This 
is of course because the number of possible rotations grows with 
misorientation angle, like the area of a sphere in rotation space. This
has the important consequence of making $\rho(\theta)$ vanish at 
$\theta=0$; this reflects not some special physics which avoids small
misorientation angles, but a simple geometrical fact that a small 
net misorientation angle demands three independent rotation angles all 
being small. Indeed, we will see that the experimental misorientation
distribution vanishes not as $\theta^2$ as would seem natural from equation
(\ref{eqn:LambdaNToRhoTheta}), but as $\theta$. We will explain this,
and the corresponding cusp in $\Xi(n)$, when we incorporate the effects
of cell division, which provides a source of new boundaries at zero 
misorientation angle.

\begin{figure}[thb]
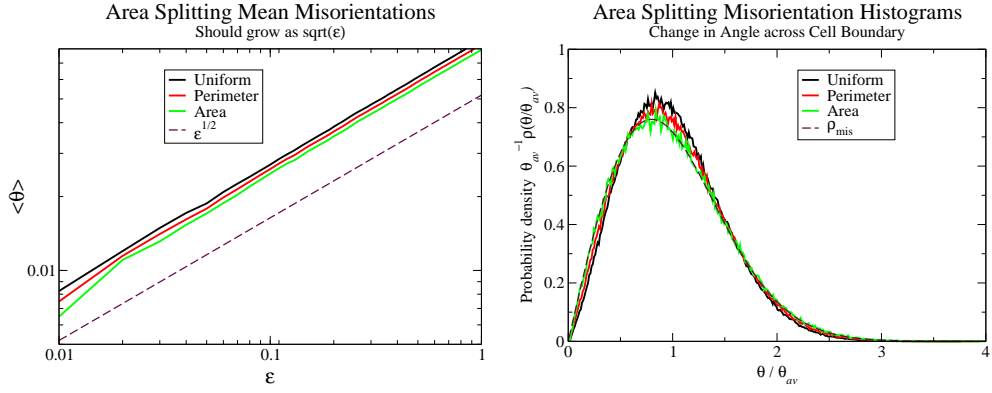

\begin{center}
\epsfxsize=2.5truein
\epsffile{MeanMisorientations.eps}
\hskip 0.1truein
\epsfxsize=2.5truein
\epsffile{Misorientation.eps}
\end{center}
\caption{{\bf Misorientation Distributions: Area Splitting.} 
The misorientation angles between cell walls, weighting each equally
(Uniform), or weighting by Perimeter or Area of the cell boundary.
LEFT: the mean misorientation angle grows as $\epsilon^{1/2}$ in our model,
not only for Area splitting (shown) but also for the other forms of the cell
splitting rate. RIGHT: the scaling form for the probability distribution
is largely independent of the way in which one weights the cells by their
size, and agrees well with the form $\rho_{\rm mis}$ of 
equation~(\protect\ref{eqn:SolutionScaling}).
}
\label{fig:Misorientations}
\end{figure}

Figure \ref{fig:Misorientations} shows the misorientations we measure for
our Area splitting model. As for the cell sizes, the misorientations
across cell walls can be averaged weighting them uniformly, by
the cell wall perimeter, or by its area.

We see that the mean misorientation angle in our model grows with 
$\epsilon^{1/2}$, as does the experimental misorientations across the 
IDB boundaries. We find similar scaling for Volume splitting, and
roughly similar for Uniform splitting dynamics; our derivation below
suggests that this scaling is generic for our rotational diffusion mechanism.
Hence our model will not provide an explanation for the 
$\theta_{\rm av}^{(GNB)} \sim \epsilon^{2/3}$ seen for the GNBs.

Since the GNBs are in practice distinguished from the IDBs by the number
of perpendicular cell walls impinging upon them (each GNB has typically
a couple of IDBs), we tested whether separating our cell boundaries into
previously-split GNB analogs and unsplit IDB analogs might lead the 
misorientations of the former
to grow more quickly on average with external strain. This did not occur in
our model: both previously split and unsplit cell walls scale in mean
misorientation with the square root of the external strain.

We can derive a simple differential equation of the time evolution of 
the distribution of misorientation matrices across incidental boundaries 
$\Xi({\bf n},\epsilon)$. Since new cell
boundaries are created at zero angle, the equation will be a diffusion 
equation with a source term:
\begin{equation}
{\partial\Xi({\bf n})\over\partial \epsilon} 
	= 2 D \nabla^2 \Xi + C \epsilon^{{1\over3\gamma}-1} \delta({\bf n}).
\label{eqn:DLambdaDeps}
\end{equation}
where the gradients on the right are with respect to $\bf n$, and 
$\delta({\bf n})$ is a three-dimensional Dirac delta function (infinite at
zero, zero elsewhere, integral equal to one). The misorientation matrices
diffuse with coefficient $2D$ because the two cells on either side are 
each diffusing with diffusion coefficient $D$.

The first term in equation (\ref{eqn:DLambdaDeps}) represents the diffusion, or
random walk, in rotation space. 
The second term in equation (\ref{eqn:DLambdaDeps}) represents the creation
of new cell boundaries which divide the old ones. 
Because we choose to weight our misorientation
angle density according to the boundary area, this division does not change
the cell boundary density $\Xi({\bf n})$ except at ${\bf n}=0$.

The new boundary area shows up in our distribution at zero misorientation
angle $\delta({\bf n})$. To derive the amount of new boundary area which is
created per unit strain, we use a simple scaling argument.
Note that the cell sizes scale as
$L \sim \epsilon^{-1/3\gamma}$. The total boundary area $A_{tot}$ will scale as the
number of boundaries $1/L^3$ times the area per boundary $L^2$, hence as 
$\epsilon^{1/3\gamma}$. The new boundary area needed per unit increment
of $\epsilon$ thus scales as $\epsilon^{{1\over3\gamma}-1}$,
giving the prefactor for the $\delta$-function in equation 
(\ref{eqn:DLambdaDeps}).

We now specialize to $\gamma=2/3$ corresponding to Area splitting.
If we start with no cell boundaries at $\epsilon=0$, then the solution to
equation (\ref{eqn:DLambdaDeps}) is
\begin{eqnarray}
\Xi({\bf n}) 
&=& \int_0^\epsilon du C u^{-1/2} \cr
& & \ \ \ \exp(-{\bf n}^2/ 8 D (\epsilon-u)) / (8 \pi D (\epsilon-u))^{3/2} \cr
&=& C \exp(-{\bf n}^2/8 D \epsilon) / (8 \pi D n \sqrt{\epsilon}).
\label{eqn:Solution}
\end{eqnarray}
(The cell boundaries which are formed at deformation $u$ have spread out into
a Gaussian of variance $D (\epsilon - u)$) The total boundary area $A_{tot} = 2 C
\sqrt{\epsilon}$, as desired. This leads to a prediction for the 
probability distribution of misorientation angles that yields the scaling
collapse (\ref{eqn:ScalingAngle})
\begin{eqnarray}
\rho(\theta,\epsilon) 
	&=& \theta_{av}^{-1} \rho_{\rm mis}(\theta/\theta_{av})\cr
	&=& \theta \exp(-\theta^2/8 D \epsilon)/4 D \epsilon 
\label{eqn:SolutionTheta}
\end{eqnarray}
where 
\begin{equation}
\theta_{av} = \sqrt{2 \pi D \epsilon}
\label{eqn:ThetaAverage}
\end{equation}
and
\begin{equation}
\rho_{\rm mis}(x) = (\pi x / 2) \exp(-\pi x^2/4)
\label{eqn:SolutionScaling}
\end{equation}
(As we will discuss in the appendix \ref{sec:Connection}, this happens to be the
distribution as derived by Pantleon\cite{PantleonScripta,PantleonActa}
in a model without cell refinement.)
As shown on the right in figure~(\ref{fig:Misorientations}), this scaling
form describes the simulation well: the
simple scaling argument for the source term of new boundaries (above) captured
the behavior of the stochastic simulation.

\begin{figure}[thb]
\begin{center}
\null\hskip -0.15truein
\leavevmode
\epsfxsize=9.cm
\epsffile{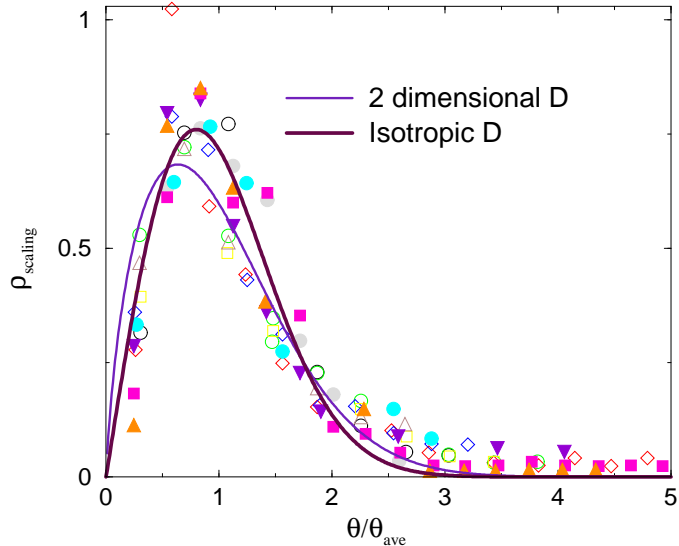}
\end{center}
\caption{{\bf Experiment vs. Theories.} 
Experimental data digitized from Hughes {\it et al.}\protect\cite{HughesPRL}.
The thick curve is $\rho_{\rm mis}$, the scaling function of equation
(\ref{eqn:SolutionScaling}). 
The thin curve is $\rho_{\rm anisotropic}$, the solution of equation
(\ref{eqn:DLambdaDepsPantleon}) with a length-independent diffusion constant
with one zero eigenvalue, representing a a highly anisotropic rotational
diffusion. 
}
\label{fig:ExperimentVsTheory}
\end{figure}

One can see from figure \ref{fig:ExperimentVsTheory} that 
the predicted form is also quite a good description of the experimental data.
This particular functional form is derived using $L \sim \epsilon^{-1/2}$,
but the particular exponent is not crucial to the analysis: the solution
for $L \sim \epsilon^{-{1\over3\gamma}}$ has this general form 
(with linear slope at $\theta=0$ and $\theta_{av}(\epsilon)\sim\sqrt{\epsilon}$)
for other $\gamma$ as well.

\section{Conclusion: Missing pieces}
\label{sec:Conclusion}

We believe our simple model is the explanation for the observed scaling
seen in cell refinement during plastic flow, despite many missing
features that are clearly relevant to the experimental behavior. In
particular, we would claim that any physical system that refines by
subdivision, and that undergoes random rotational distortions, is likely
to enter a scaling regime similar to that seen in our model. In this
section, we discuss three of the missing features and why they must be
relevant for a complete engineering description.

\noindent (1)~{\bf IDB/GNB Distinction.} The experiments show two
distinct types of cell boundaries, with different scaling exponents and
scaling functions. Our model only includes one. While we get a plausible
fit to the IDB scaling, the GNB misorientation angles grow faster than
our rotational diffusion model can reproduce. One must note that scaling
behavior in other contexts is associated with a single characteristic
length scale: our model asymptotes to a morphology which is
statistically unchanging at large strains except for a single rescaling
of length. Because the experiments show two length scales, we expect
that detailed measurements will show violations of scaling behavior,
say, when the mean GNB separation length crosses the IDB separation
length. Perhaps the fact that the two types of boundaries are roughly
evolving on perpendicular axes allows for the approximate scaling seen.

\noindent (2)~{\bf Plastic Deformation.} In our simple model, the
overall plastic deformation was ignored. In a real material under
compression the boundaries perpendicular to the strain axis will grow
closer to one another linearly in the strain even without subdivision!
Apparently\cite{PantleonPrivateCommunication} under the
external strain the GNB's slowly rotate towards this axis, while the
IDB's remain roughly aligned parallel to the axis of compression. The
external strain, after this rotation is complete, will act to separate
the IDB's roughly as the square root of the strain $\epsilon^{1/2}$. To
get the observed refinement $\epsilon^{-1/2}$ the subdivisions must
happen even faster than described by the area splitting law we focused
on in the text: perimeter splitting, which in the undeformed coordinates
leads to a length which scales as $\epsilon^{-1}$ would work. This begs
the question of why the scaling should begin to be observed even before
the rotations are complete.

Pantleon\cite{PantleonPrivateCommunication} suggests that this might be an
explanation for the observed $\epsilon^{-2/3}$ scaling of the separation
between GNBs. At short times, before the rotation is complete, the
lengths vary as $\epsilon^{-1/2}$, and at long times (with the added
effects of compression) they will vary faster, perhaps leading to a
reasonable fit with the larger exponent.

\noindent (3)~{\bf Origins of the cell splitting, rotational diffusion.}
In our model, the cell splitting and rotational diffusion are given as
part of the dynamics: we do not address what the physical mechanisms are
that produce them.

There are various proposed mechanisms for getting the cell sizes to
shrink. Obviously the cell walls cannot just move inward: the cell wall
velocities would grow linearly in the system size. One could imagine a
crinkling of existing cell walls (the inverse of the coarsening process
seen in spinodal decomposition), or nucleating new cells at junctions of
existing cell walls: neither picture is compatible with our analysis,
and both involve cell wall motion which is resisted by pinned sessile
dislocation junctions\cite{PrinzArgon}. Subdivision as we've used it
could arise from collisions between dislocations as they traverse the
cell, although simple calculations suggest that the expected collision
rates are too small\cite{ArgonHaasen,PrinzArgon} and don't scale
correctly for our theory\cite{EarlyDraft}. Cell splitting due to
inhomogeneous stresses induced by neighboring cells\cite{Dawson} seem to
us the most natural and likely mechanism. A corresponding microscopic
picture would involve regions in the inhomogeneously stressed cell where
the dislocations slow down, or where they are more easily pinned by
obstacles or other dislocations, leading to the formation of a new cell
wall.

The mechanisms driving the rotations of the crystalline axes of the
cells are less well understood. The crystalline axes can rotate both
directly through the rotation of the material in the cell, and
indirectly because of the flux of dislocations mediated by the plastic
deformation. This latter effect is well studied on larger scales in the
field of texture evolution, where the plastic deformation of a
polycrystal often leads to a gradual alignment of their crystalline
orientations. Within a single crystalline grain, this texture evolution
will on average rotate all the cells together: because the relative
angles between cells is small, they will largely rotate in the same
direction. One should note that the traditional explanation for the
origin of the GNBs indicates that their misorientation angles could well
have a overall mean drift in addition to the random diffusion. It is
said that GNBs are observed to have rotational misorientations that
alternate in sign, because they separate regions with differing active 
slip systems \cite{HughesPrivateCommunication};
the combination of the differing plastic strains and rotations in
neighboring pairs of cells can equal the net imposed plastic
deformation. In this picture, the net rotation angles across the GNB's
(equation~\ref{eqn:OrientationalDiffusion}) should have a mean drift
term in addition to the diffusion term.


\noindent {\bf Acknowledgments}. This work was supported by the Digital
Material project NSF KDI-9873214, the ITR/ASP project NSF ACI0085969,
and the Harvard Society of Fellows. We thank Paul Dawson and Ali Argon
for useful conversations, Karin Dahmen for help with the manuscript,
and Wolfgang Pantleon and Darcy Hughes for constructive criticism.

\section{Appendix: Connections with Stochastic Dislocation Theories}
\label{sec:Connection}

The distribution for the misorientation angle for the simple rotational
diffusion model, $\rho_{\rm mis}(x)$, happens to have the same form
as one derived by Pantleon\cite{PantleonScripta,PantleonActa} without
considering cell
refinement, and by assuming that the noise in the cell orientations was
due to random, uncorrelated fluctuations in the dislocation flux. It behooves
us, therefore, to discuss how Pantleon's work can be interpreted
in our context.

Pantleon's theory, also suggested by Nabarro\cite{Nabarro} and Argon and
Haasen\cite{ArgonHaasen}, is that the stochastic noise in the flux of
dislocation from either side of the cell boundary leads to randomness
in the evolution of the cell boundary angles. Each dislocation passing
through a cell, say, may shift the top plane of atoms by a distance $b$
with respect to the bottom plane, where $b$ is the Burgers vector of the
dislocation (roughly the lattice constant). The crystalline axes within
a cell of characteristic height $L$ will rotate due to one dislocation
an amount proportional to $b/L$.  Under a strain increment 
$\Delta \epsilon$, a cell of characteristic height $L$ must have 
$N= L \Delta \epsilon / b$ dislocations impinging on the side cell boundary.
A roughly equal and opposite average flux will impinge on the cell boundary
from the cell on the other side. If the dislocations move independently
(which we will argue does not occur), then one expects that there will
be a net residue after the strain increment of roughly 
$\sqrt{N} = \sqrt{L \Delta \epsilon / b}$.  Hence
the predicted drift in angle after a strain increment of $\Delta \epsilon$
is $\Delta\theta = \sqrt{N} b/L = \sqrt{b \Delta \epsilon/L}$. The diffusion
constant is given by $D(L) \sim (\Delta\theta)^2 / \Delta\epsilon \sim b/L$.
Because a single dislocation produces a larger net rotation for smaller
cell, the stepsize in the random walk in rotation space becomes larger
as our cells get smaller.

Pantleon and Hansen\cite{PantleonHansenActa} consider three cases, where
one, two, and three slip systems are activated. The geometry of a given
cell -- the direction and strength of the applied shear with
respect to the crystalline axes -- will determine
what types of dislocations are allowed to pass through the cell. If
only one slip system is active, the rotation of the cell will be confined
to a single axis. In our formulation, the diffusion constant in rotation 
space will be anisotropic: in this case it will be a rank one tensor
(a $3 \times 3$ matrix with only
one non-zero eigenvalue). Two slip systems will give a rank 2 tensor, with
one zero eigenvalue. Three slip systems will allow the cell to diffuse
in any direction, but even so the diffusion constant $D$ will in general
be anisotropic.

To make contact with Pantleon, we consider a more general evolution
law for the misorientation in area splitting:

\begin{equation}
{\partial\Xi(n)\over\partial \epsilon}
= D_{ij}(L) \nabla_i \nabla_j \Xi
+ C \epsilon^{1/2} \delta(n).
\label{eqn:DLambdaDepsPantleon}
\end{equation}

Here we have allowed for an anisotropic, cell-size-dependent diffusion constant
by introducing the symmetric tensor $D_{ij}(L)$ depending on the current
cell size $L$.  The symmetric diffusion tensor $D_{ij}$ will vary with 
the geometry of the particular cell boundary: it can depend both on the
relative orientation of the crystal and the strain tensor with respect to
the plane of the cell boundary. We'll first consider the anisotropy while
ignoring the refinement ($D_{ij}(L)$ independent of $L$), in analogy to 
previous work, and then incorporate the refinement ($D_{ij}(L)$ proportional
to $1/L \sim \epsilon^{1/2}$ as argued above).

Pantleon\cite{PantleonScripta,PantleonActa,PantleonHansenActa} in most
of his analysis ignores cell refinement. If we make $D_{ij}$ independent
of $L$ and set $C=0$ in equation (\ref{eqn:DLambdaDepsPantleon}), we
get an anisotropic diffusion equation whose solution (assuming a narrow
initial distribution of misorientations) is an anisotropic Gaussian with 
variances given by the inverse eigenvalues of $D$, with the 
experimentally observed scaling $\theta_{av}\sim\epsilon^{1/2}$. 
If we assume $D$ has one non-zero eigenvalue, we get the Gaussian
distribution derived by Pantleon and Hansen for one active slip system.
If we assume $D$ has two equal, non-zero eigenvalues, we get the
Rayleigh distribution that they find for two perpendicular systems of edge
dislocations (which, coincidentally, is the same distribution that
we found above (\ref{eqn:SolutionScaling}) with an isotropic $D$
and a source term). If we assume $D$ is isotropic, we get the Maxwell
distribution they find for three perpendicular systems of dislocations.

What happens to the solution of equation (\ref{eqn:DLambdaDepsPantleon})
when we incorporate cell refinement? Pantleon\cite{PantleonActa} notices
that non-constant cell size must change the scaling of average angle with
strain. By setting $D(L)\sim\epsilon^{1/2}$ in accordance with reference
\cite{HughesActa}, and changing variables
to $\tau=\epsilon^{3/2}$ in equation (\ref{eqn:DLambdaDepsPantleon}),
we can map it into a problem quite similar
to equation (\ref{eqn:DLambdaDeps}) except that the source term is
of magnitude proportional to $\tau^{-2/3}$. The solution to
this equation has a shape quite similar to that shown in figure
(\ref{fig:ExperimentVsTheory}), but
with an average angle $\theta_{av} \sim \sqrt{\tau} \sim \epsilon^{3/4}$.
This yields a $3/4$ power of the strain incompatible with the scaling observed
for the incidental cell boundaries.  Pantleon is aware of 
this\cite{PantleonActa}: in an analogous
calculation he gets $\theta_{av}\sim\epsilon^{0.72}$.

(If we abandon dislocation noise as the origin of the rotational diffusion,
and take the crystalline lattice orientations as our basic variables, then
there are no microscopic lengths remaining in the problem. The diffusion
constant then naturally depends only on the local geometry of the cell
boundary, and hence is independent of the length scale.)

The key to the discrepancy is of course that the dislocations motion is
not uncorrelated: they must be moving in a collective manner.
The interaction energy between dislocations is large,
and diverges logarithmically with distance, reflecting the infinite stiffness
of a crystal to gradients in the axes of rotation. In early stages of 
hardening, one might plausibly argue that the dislocations are sufficiently far
apart that their interaction forces do not dominate. But in the regime
studied here, the fact 
that the dislocations organize into boundaries (avoiding rotational distortions 
within the cells) is a clear signal that it is no longer sensible to treat their
evolution independently. If the top half of a cell boundary received more
dislocations from the right-hand cell than the bottom half, this would
produce an enormous bending force on the cell. Such an event could only happen
for reasonable energy cost if this bending were screened by the division of
the cell by a new boundary.

\begin{figure}[thb]
\begin{center}
\leavevmode
\epsfxsize=8.cm
\epsffile{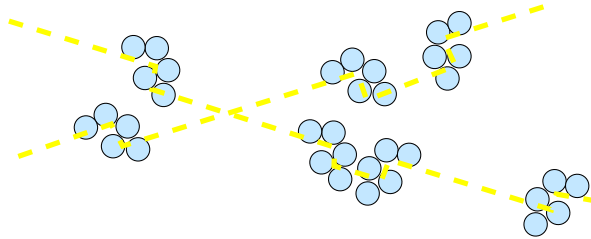}
\end{center}
\caption{{\bf Correlated atomic rearrangements associated with simple 
dislocation motion.} 
}
\label{fig:KinkZipper}
\end{figure}

The stochastic dislocation flux models will be generally applicable
whenever the correlations between their motions vanish at distances 
comparable to the cell sizes. Our model is correct in the other limit:
our cells only rotate as units. Thus our model is appropriate for systems
where the dislocation motions are strongly correlated on the scale of the
cells -- so that their motion can always be described as mediating 
overall rotations of each cell. Both descriptions are only a starting
point for a complete theory.

Consider an analogy: dislocation motion modeled as rearrangements of atoms,
versus as the evolution of a continuum curve.
In the atomic description, one would identify
characteristic atomic motions (say, kink diffusion events in the case of
semiconductors), at rare sites scattered through the crystal. These sites
would be strongly correlated (lying
along the dislocations); their dynamics would produce unusual zipper motions 
(kinks diffusing along dislocations).  We expect similar correlations
to arise in the dislocation motions mediating cell boundary evolution:
the dislocations will form correlated dances to keep rotational gradients from 
entering the cells. 
On the other hand, treating the dislocation motion as the evolution of a
continuous curve makes it difficult to incorporate the anisotropic
dynamics and lattice pinning effects -- our model so far has ignored the 
corresponding crystalline and shear anisotropies in the evolution of cell 
structure.

We can incorporate some of this asymmetry by hand into our model.
Pantleon and Hansen\cite{PantleonHansenActa}
point out that the individual cell boundaries have
rather low symmetries. The diffusion tensor in equation 
(\ref{eqn:DLambdaDepsPantleon}) describes the evolution of that subset
of cell boundaries with a particular cell boundary orientation and
(average) crystal lattice orientation with respect to the external
shear. There is no reason that for a low-symmetry geometry that the
diffusion constant will be isotropic, geometry dependent, or material
dependent: the general evolution law is (\ref{eqn:DLambdaDepsPantleon})
with $D_{ij}$ independent of $L$. One must solve for the distribution at 
fixed geometry and then average over geometries to predict the experimental
distribution (as also discussed in section 5.2 of reference 
\cite{PantleonHansenActa}). If we assume a strongly anisotropic
rotational diffusion tensor with one zero eigenvalue and the other two equal
(corresponding roughly to Pantleon's analysis with two active slip systems)
we can solve equation (\ref{eqn:DLambdaDepsPantleon}) to find a scaling
collapse (\ref{eqn:ScalingAngle}) with scaling function
\begin{equation}
\rho_{\rm 2dim}(x) = (\pi^3 x/64) \exp(-\pi^3 x^2/128) K_0(\pi^3 x^2/128)
\label{eqn:2dim}
\end{equation}
where $K_0$ is the modified Bessel function of the second kind. This scaling
function is shown in figure (\ref{fig:ExperimentVsTheory}).

It is important also to note that effects that seem clearly related to 
cell boundary formation have been seen in finite-element simulations of
single-crystal plasticity by Mika and Dawson {\it et al.}\cite{Dawson}. 
Their system consisted of several crystalline grains with differing
orientations, subject to an external shear.  The inhomogeneous strains
within the grains led to the formation of subgrain structures very
similar to cells. The distribution of cell boundaries in their simulation 
was also found to scale, with strain dependence and functional form
similar to that seen for geometrically necessary boundaries (that is,
scaling with the $2/3$ power of strain, and not with the $1/2$ power
seen for the IDBs).
Hence Dawson {\it et al.} find cells in a simulation totally without 
dislocations.

\end{document}